\documentclass[a4paper,11pt]{article}
\usepackage{jcappub} 

\usepackage{booktabs}
\usepackage[dvipsnames]{xcolor}
\usepackage{colortbl}
  \newcommand{\sampled}{\cellcolor{gray!20}}
  
\usepackage{lineno}

\newcommand{\vtheta}{\boldsymbol{\theta}}
\newcommand{\As}{A_\mathrm{s}}
\newcommand{\ns}{n_\mathrm{s}}
\newcommand{\Omegam}{\Omega_\mathrm{m}}
\newcommand{\Omegac}{\Omega_\mathrm{c}}
\newcommand{\Omegab}{\Omega_\mathrm{b}}
\newcommand{\OmegaL}{\Omega_\Lambda}
\newcommand{\zetaUV}{\zeta_\mathrm{UV}}
\newcommand{\HI}{\mathrm{HI}}
\newcommand{\He}{\mathrm{He}}
\newcommand{\HeII}{\mathrm{HeII}}
\newcommand{\e}{\mathrm{e}}
\newcommand{\re}{\mathrm{re}}
\newcommand{\reio}{\mathrm{reio}}
\newcommand{\rec}{\mathrm{rec}}
\newcommand{\ap}{\alpha}
\newcommand{\tilt}{\beta}
\newcommand{\ar}{\tilde{a}}
\newcommand{\gomp}{\mathrm{gomp}}

\arxivnumber{2405.13680} 
\title{\boldmath Five parameters are all you need \\(in $\Lambda$CDM)}

\author[a,*]{Paulo Montero-Camacho }
\author[a,*]{Yin Li }
\author[b,c]{Miles Cranmer}
\affiliation[a]{Department of Strategic and Advanced Interdisciplinary Research, Peng Cheng Laboratory,\\ Shenzhen, Guangdong 518000, China}
\affiliation[b]{Institute of Astronomy, University of Cambridge,\\ Madingley Road, Cambridge, CB3 0WA, UK}
\affiliation[c]{Department of Applied Mathematics and Theoretical Physics, University of
Cambridge,\\ Wilberforce Road, Cambridge, CB3 0WA, UK}
\affiliation[*]{\emph{With equal contribution}}

\emailAdd{pmontero@pcl.ac.cn, eelregit@gmail.com}

\abstract{The standard cosmological model successfully describes the Universe with six parameters. One of them, the optical depth to reionization $\tau_\reio$, empirically depicts the scatterings of Cosmic Microwave Background (CMB) photons off free electrons, as the intergalactic medium transitions from neutral to ionized. $\tau_\reio$ depends on the neutral hydrogen fraction $x_\HI$, which ultimately depends on cosmology. We present a novel method to establish such a missing link between physical parameters and reionization timeline using symbolic regression. We discover the timeline has a universal shape well described by the Gompertz mortality law, applicable to any cosmology within our simulated data. Unlike the conventional tanh prescription, our model is asymmetric in time and a good fit to astrophysical constraints on $x_\HI$. By combining CMB with astrophysical data and marginalizing over astrophysics, we treat $\tau_\reio$ as a derived parameter, tightening its constraint to $<3\%$. This approach reduces the error on the amplitude of the primordial fluctuations by a factor of 2.3 compared to Planck's PR3 constraint and provides a commanding constraint on the ionization efficiency $\zetaUV = 26.9^{+2.1}_{-2.5}$. We expect further improvements in the near term as reionization constraints increase and our understanding of reionization advances.}

\begin{document}
\maketitle
\flushbottom

\section{Introduction}
\label{sec:intro}
The $\Lambda$CDM cosmological model has proven extremely effective in predicting the evolution of our Universe, relying on only six parameters \cite{Planck2020a}. In particular, it explains the transition from a predominantly neutral state in the early stages to the familiar ionized intergalactic medium (IGM) observed in our relatively nearby surroundings. This transition is known as cosmic reionization. Despite a comprehensive understanding of the astrophysical principles governing this transition, uncertainties persist regarding its precise timeline \cite{Jin2023}. The advent of the James Webb Space Telescope (JWST) \cite{Gardner2006} represents a pivotal moment, substantially bolstering our ability to directly constrain the evolution of the neutral hydrogen fraction $x_\HI$. This progress is being driven by JWST's enhanced detection capabilities, enabling the observation of high-redshift quasars \cite{Eilers2023} and high-redshift galaxies \cite{Adams2023, Bradley2023, Donnan2023,
Ning2024}.

Reionization leads to scattering of Cosmic Microwave Background (CMB)
photons by free electrons, disrupting the CMB angular power spectra
$C_\ell$.
This scattering suppresses the signal at scales smaller than the Hubble
scale at reionization (approximately $\ell>10$) \cite{Planck2020b} due
to the optical depth $\tau_\reio$.
Additionally, it introduces a new signal in the polarization of CMB
photons at large angular scales \cite{Planck2020a}, that is $\propto
\tau_\reio$ in $C^{TE}_\ell$, the cross-correlation of the $E$-mode
polarization with the temperature (intensity), and is $\propto
\tau_\reio^2$ in $C^{EE}_\ell$, the $E$-mode polarization angular auto
power spectrum.
Consequently, heightened sensitivity to CMB polarization becomes crucial
for mitigating the degeneracy between $\tau_\reio$ and other
cosmological parameters, particularly $\As$, the amplitude of the
primordial scalar power spectrum, and $r$, the ratio of tensor-to-scalar
modes \cite{Natale2020}.

\begin{figure*}[!htb]
\centering
\includegraphics[width=0.6\textwidth]{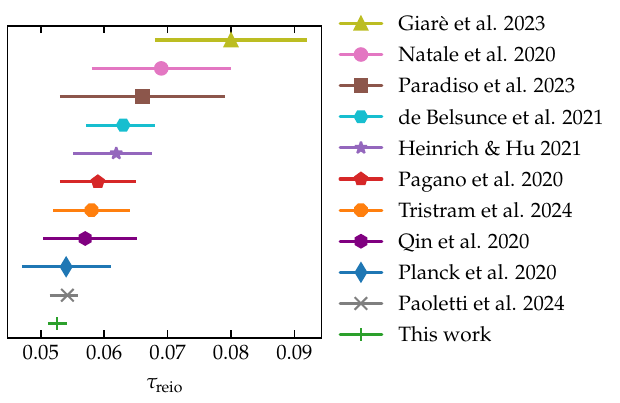}
\caption{Current constraints on the optical depth to reionization $\tau_\reio$ from Cosmic Microwave Background (CMB) data. The error bars indicate the 1$\sigma$ uncertainties. Various analyses may employ distinct data sets or vary in the parameters considered. For instance, the inclusion of astrophysical data \cite{Qin2020, Paoletti2024} (cross and purple hexagon), WMAP data \cite{Natale2020,Paradiso2023} (circle and square), or ACT in combination with other external data sets \cite{Giare2023} (triangle), expanded sky coverage \cite{Paradiso2023} (square), incorporation of high-$\ell$ data \cite{Pagano2020, Planck2020a, HeinrichHu2021, Giare2023, Tristram2024} (pentagon, diamond, star, triangle, and octagon), joint low-$\ell$ TT and EE analysis \cite{deBelsunce2021} (cyan hexagon), marginalization over small set of strongly correlated parameters \cite{Natale2020} (circle), and the implementation of an end-to-end Bayesian framework that marginalizes over astrophysics and instrumental systematics \cite{Paradiso2023} (square).}
\label{fig:tau}
\end{figure*}

Low-$\ell$ polarization data is crucial to determine $\tau_\reio$;
however, the measurement of such a weak signal ($\sim 10^{-2} \mu$K$^2$)
demands superb systematic and foreground control \cite{Planck2020b}.
Furthermore, anomalous measurements in $C^{TE}_\ell$ at low multipoles
\cite{Planck2020a} could indicate concerns to the cosmological
interpretations at these angular scales.
Ultimately, this challenging measurement may require adopting a
comprehensive Bayesian framework to jointly consider cosmology,
astrophysics, and instrument systematics \cite{Paradiso2023}.
Figure~\ref{fig:tau} illustrates current representative constraints on
$\tau_\reio$.

Given the challenges posed by $\tau_\reio$ in CMB analyses and the
anticipated advancements in constraining the reionization
timeline \cite{Montero2021, Hera2022}, now is an opportune moment to
reassess its role.
Theoretically, cosmic reionization is uniquely determined by cosmology,
i.e., $x_\HI(z)$ is fully determined by the other five cosmological
parameters.
However, incomplete understanding of reionization obscures this mapping,
necessitating the additional empirical parameter $\tau_\reio$ in CMB
analyses.
Since the inclusion of $\tau_\reio$ became a standard practice, our
understanding of the astrophysical processes governing reionization has
significantly improved \cite{Gnedin2022, Kannan2022,Murray2020, Fan2023}
and ongoing and forthcoming observations promise to reduce inherent
modeling uncertainties.

Motivated by these developments, we use symbolic regression (SR)
\cite{Cranmer2023, Graham2013} to construct a mapping between cosmology,
astrophysics, and reionization timeline, aiming to demote $\tau_\reio$
from an independent to a derived cosmological parameter and
simultaneously tightening constraints in cosmological and astrophysical
parameters.
This mapping can also shed light on reionization astrophysics and aid
ongoing efforts in parametrizing reionization models \cite{Trac2018,
Trac2022, Paoletti2024} by including the cosmological dependence of
$x_\HI$.

\begin{figure*}[!htb]
\centering
\includegraphics[width=0.9\linewidth]{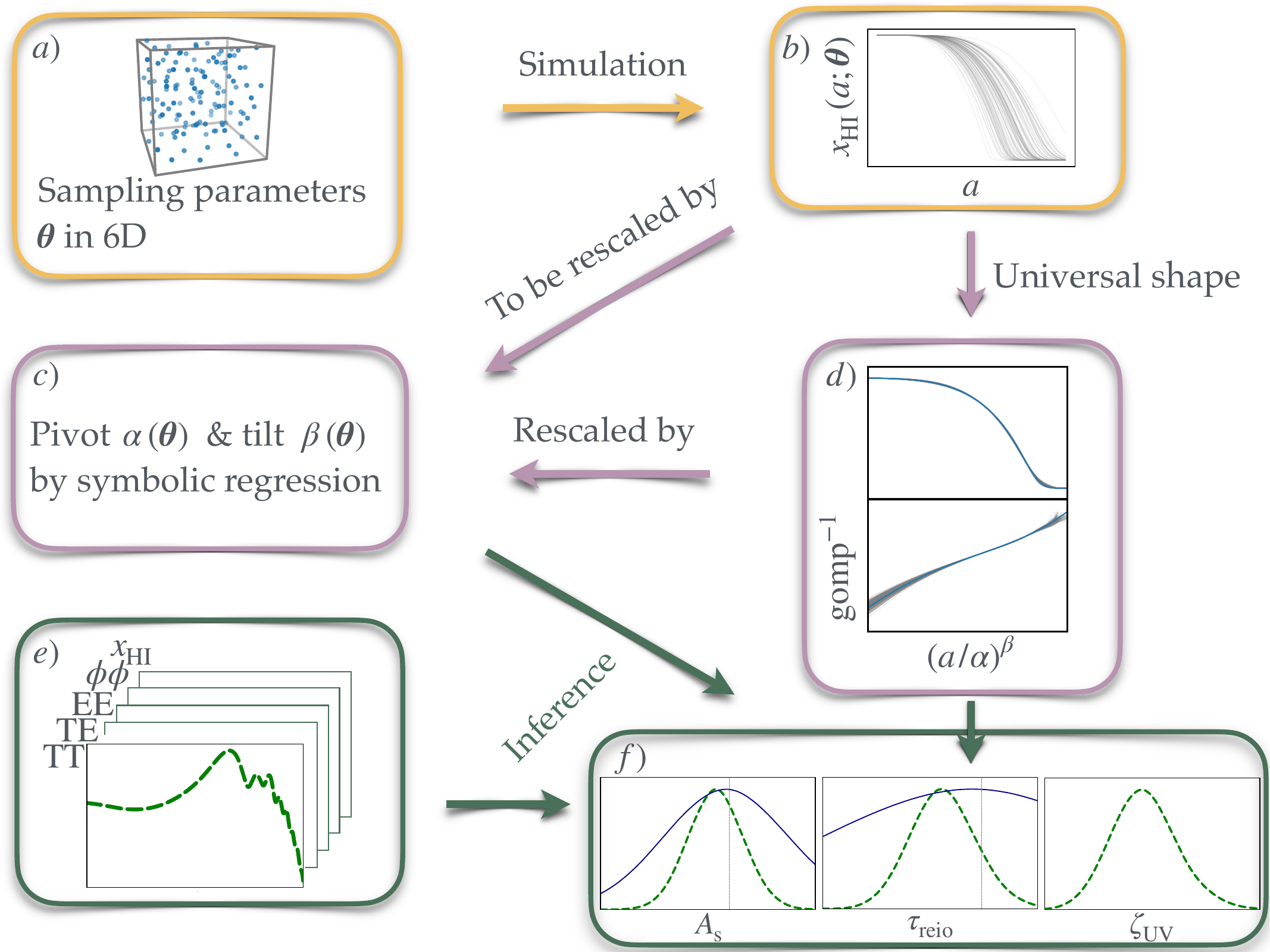}
\caption{Strategy to demote $\tau_\reio$ to derived parameter.
(\textbf{a}) Sobol sampling of $\vtheta$ comprising 5 cosmological and 1
astrophysical (nuisance) parameters (see Figure \ref{fig:sobol}).
(\textbf{b}) Simulated $x_\HI$ timelines as a function of $\vtheta$ and
scale factor $a$.
(\textbf{c}) With symbolic regression, we optimize the mapping from
$\vtheta$ to the rescaling parameters that bring the universality.
(\textbf{d}) We model the universal shape (upper panel) as a composition of
the Gompertz function and a low-degree polynomial (lower panel).
(\textbf{e}) Planck CMB data and $x_\HI$ data we analyze.
(\textbf{f}) We infer the parameter constraints using Monte Carlo Markov
Chain (MCMC).}
\label{fig:big}
\end{figure*}

Here, we present a universality in the neutral hydrogen time evolution
related via a power-law transform, and derive through SR the
dependence of power-law parameters on cosmology and astrophysics from
simulated \textsc{21cmFAST} \cite{MesingerEtAl2011, Murray2020}
reionization histories.
We integrate this universally shaped reionization timeline into
\textsc{CLASS} \cite{Blas2011}, a popular Boltzmann solver for CMB
analyses.
We then evaluate the modified \textsc{CLASS} alongside \textsc{Cobaya}
\cite{Torrado2020}, a speed-aware sampler \cite{Lewis2002,
Lewis2013}, showcasing its ability to
recover parameter constraints from CMB data, including `TTTEEE' +
lensing likelihoods \cite{Planck2020c, Planck2020d} (see Figure \ref{fig:unleashed_gomp}).
Finally, combining CMB and astrophysical data, we marginalize over
reionization astrophysics to compute $\tau_\reio$ as a derived parameter
using SR, quantifying the gains compared to sampling over $\tau_\reio$
utilizing the conventional $\tanh$ model \cite{Lewis2008}.
We summarize our strategy in Figure \ref{fig:big}.

\section{Methodology}
\label{sec:method}
As illustrated in Figure \ref{fig:big}, our strategy to demote $\tau_\reio$ into a derived parameter requires simulations of the evolution of neutral hydrogen fraction, identification of an appropriate universal shape (or template), and establishment of the connection between cosmology and reionization astrophysics. Here we will introduce the components of this plan.

\subsection{Simulations}
\label{sec:sims}
\begin{figure}[tb]
\centering
\includegraphics[width=0.7\textwidth]{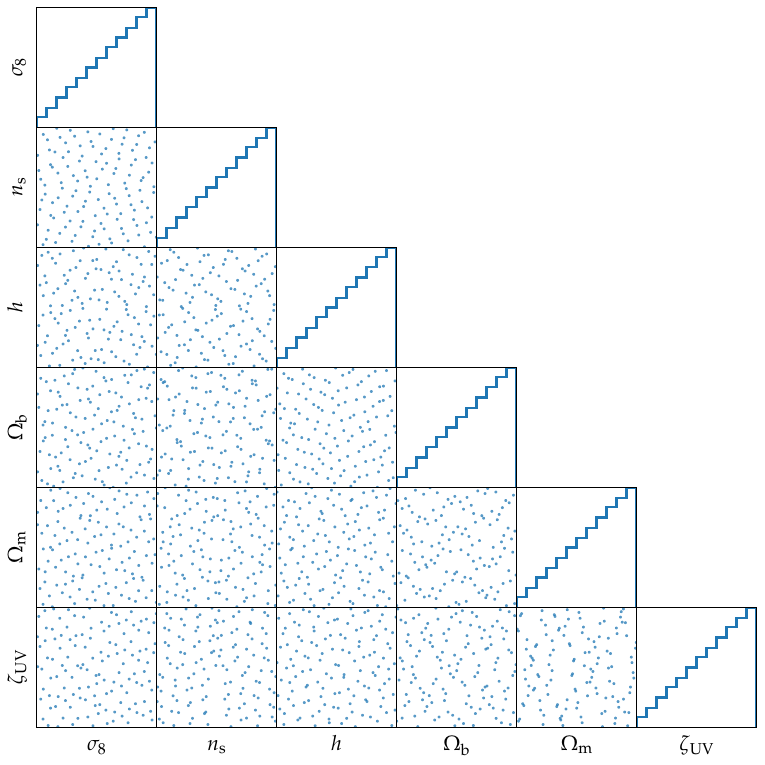}
\caption{\textbf{\boldmath Sobol design for our $x_\HI$ edge and core
samples},
in 6D parameter space of $\sigma_8$, $\ns$, $h$, $\Omegab$, $\Omegam$,
and $\zetaUV$.
Each of the 128 points in the lower triangular panels corresponds to the
2D projection of one core and one edge \texttt{21cmFAST} runs, while the
diagonal panels show the 1D cumulative histograms for each parameter.
These 1D and 2D projections demonstrate the uniformity of our sampling
of the parameter space within the edge prior range of Eq.~(\ref{eq:prior}),
as well as in the core area surrounding the Planck PR3 constraint.}
\label{fig:sobol}
\end{figure}

To establish the universality of our proposed Gompertzian model for the
neutral hydrogen reionization and its relationship with the cosmological
parameters, we conducted 256 \texttt{21cmFASTv3} simulations to generate
the corresponding $x_\HI(z; \vtheta)$ profiles.
Our parameter space include $\vtheta = \{\sigma_8, \ns, h, \Omegab,
\Omegam, \zetaUV\}$, comprising five cosmological and one astrophysical
parameters.
The selection of $\sigma_8$ instead of $\As$ is imposed by the input
requirements of \texttt{21cmFAST}\footnote{Alternatively, one could use
\texttt{21cmFirstCLASS} \cite{Flitter2024} to avoid $\sigma_8$.}.
We first use a scrambled Sobol sequence \cite{Sobol1967, Owen1998} of
length 128 to sample quasi-uniformly within the following
$\vtheta$-ranges:
\begin{alignat}{3}
\label{eq:prior}
\sigma_8 &\in (0.74, 0.90), &\quad
\ns &\in (0.92, 1.00), &\quad
h &\in (0.61, 0.73), \nonumber\\
\Omegab &\in (0.04, 0.06), &\quad
\Omegam &\in (0.24, 0.40), &\quad
\zetaUV &\in (20, 35).
\end{alignat}
Their 1D and 2D projections in Figure \ref{fig:sobol} illustrate the sample
uniformity in parameter space.
Because most of the 6D hypercube volume lies near its surface, we name
the above 128 simulations the \emph{edge} samples and let the other 128
simulations sample its \emph{core} within the 5$\sigma$ range of Planck
PR3 constraint \cite{Planck2020a}, reusing the same Sobol design in
Figure \ref{fig:sobol}.
This helps to improve the accuracy of symbolic regression where the
final posterior mass are expected to lie.

Our \texttt{21cmFAST} simulations have a 300 comoving Mpc box size and
$768^3$ ($256^3$) cells for the matter (HI) field.
We maintain most options in their default values and extract the
neutral hydrogen fraction using \texttt{lightcone.global\_xH}.

The ionization efficiency $\zetaUV$ governs the ability of ultraviolet
photons to escape their parent galaxies and ionize the IGM. An increase
in $\zetaUV$ leads to an earlier completion of HI reionization.
Due to significant uncertainties surrounding $\zetaUV$, we opted to use
a constant value in each simulation.
Note that this choice was made for simplicity, and a more realistic
assumption could be that $\zetaUV$ is a function of halo mass
\cite{Park2019} or redshift. Appendix \ref{app:halo} demonstrates that the universality holds in the case of a mass-dependent ionization efficiency. Ref. \cite{2026arXiv260413423M} explores the impact of a more robust reionization model with added nuisance parameters to account for X-ray preheating and to capture the difficulty of forming star-forming galaxies.

\subsection{Universality: Shape \& modeling}
\label{sec:uni}
\begin{figure}[tb]
\centering
\includegraphics[width=0.7\linewidth]{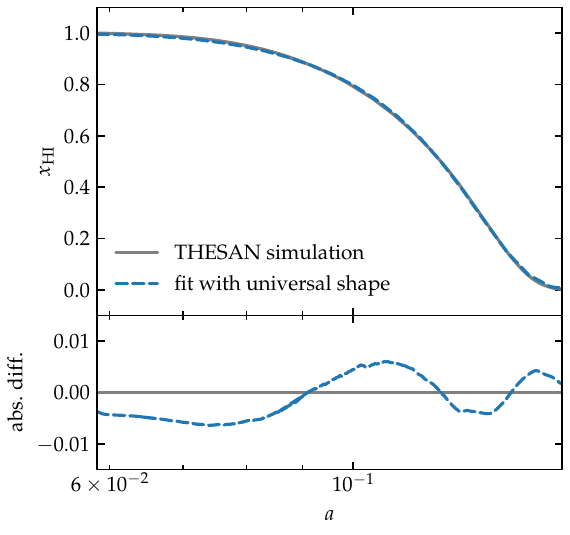}
\caption{\textbf{Universal shape of $x_\HI$ in THESAN simulation.}
\emph{(Top)} Simulated timeline from the THESAN1 simulation (gray line)
exhibits the same universality, as it can be accurately fitted by the
same analytic shape in Eqs.~(\ref{eq:uni}) and (\ref{eq:poly}), with the same coefficients
$\boldsymbol{c}$, the same power-law transformation $\tilde{a} = (a /
\ap)^\tilt$, and only $\{\ap, \tilt\}$ as the free parameters.
The blue dashed line is the fitted analytic shape model.
\emph{(Bottom)} The absolute difference between the universal shape and
the simulated data.}
\label{fig:thesan}
\end{figure}

We construct the universal $x_\HI$ shape using 256 Sobol samples
of \textsc{21cmFAST} simulations, varying 5 cosmological parameters
and the astrophysical ionizing efficiency $\zetaUV$.
The latter modulates the timing of reionization by regulating the
abundance of photons that escape into the IGM (see \S\ref{sec:sims} 
and Figure \ref{fig:sobol}).
All $x_\HI(a)$ profiles share the same shape, with differences between
scenarios being mere translations and rescalings in logarithmic scale
factor $\ln a$.
Reionization causes $x_\HI$ to reduce from near 1 to effectively 0 via a
sigmoid transition.
The standard $\tanh$ function is symmetric in nature and not flexible
enough to provide the early start and rapid completion suggested by
reionization simulations \cite{Trac2018, Doussot2019} and favored by
astrophysical data (see Figure \ref{fig:history}).
In contrast, the SR-inspired Gompertzian curve, an asymmetric sigmoid
function often used to analyze age-dependent human mortality
\cite{Gompertz1825}, proves a good model for the survival of neutral
hydrogen too.
Its expected accelerated increase in mortality with age resembles the
expectation for the percolation of ionized hydrogen bubbles during the
end stages of reionization.
Remarkably, this universality works beyond the semi-numerical
\textsc{21cmFAST} predictions and even for the state-of-the-art THESAN
simulations \cite{Kannan2022} (see Figure \ref{fig:thesan}), where accurate radiation transport and realistic
galaxy formation physics are included.

We discovered the universality in the shape of $x_\HI$ timelines before
attempting to build an  analytic model for it.
Because $x_\HI$ varies monotonically between 1 and 0, we can view it as
a cumulative probability distribution (CDF) and derive its probability density function (PDF), with which we can weigh the logarithmic scale
factor $\ln a$ to compute its mean and standard deviations.
It was immediately obvious to us that the $x_\HI$'s had a common shape
to percent level, after translation by their means and rescaling by
their standard deviations.
Therefore, cosmology and astrophysics (e.g., $\zetaUV$) only impacts the translation and
rescaling parameters of each timeline, not its shape.
However, given our broad parameter range in Eq.~(\ref{eq:prior}), some
$x_\HI$'s have not reached 0 by the end of simulations, resulting in
imperfect transformations hurting the universality.

\begin{figure}[tb]
\centering
\includegraphics[width=0.6\linewidth]{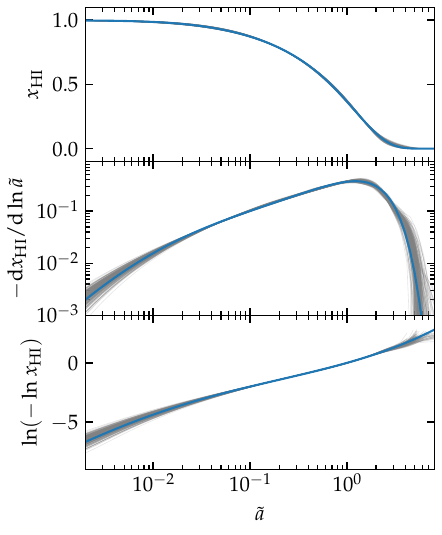}
\caption{\textbf{\boldmath Universal shape of $x_\HI$.}
\emph{(Top)} 256 simulated $x_\HI$ timelines (thin light gray lines)
exhibit universality, after power-law transformations $\ar =
(a/\ap)^\tilt$.
The blue curves in all panels show our fitted analytic shape model, a
composition of the Gompertzian curve with a 5th-degree polynomial in Eqs.~(\ref{eq:uni}) and (\ref{eq:poly}).
\emph{(Middle)} Time derivative of $x_\HI$, can be interpreted as a PDF
if we view $x_\HI$ itself as the CDF.
We first discovered the universality by translating and rescaling each
$x_\HI$ using the mean and variance of its PDF, though now switch to the
better approach that jointly fits the global shape and individual
power-law parameters.
We use the latter as target of symbolic regression.
\emph{(Bottom)} Timelines transformed by the inverse of Gompertzian
function, modeled in blue curve with a 5th-degree polynomial in
Eq.~(\ref{eq:poly}).}
\label{fig:shape}
\end{figure}

To address this, we construct flexible models for the universal shape,
and fit it jointly with individual transformation parameters of each
$x_\HI$ timeline.
We compose the Gompertz function $\gomp$, defined in Eq.~(\ref{eq:uni}), with
a low-degree polynomial $P_m$, where $m = 1, 3, 5, 7$ progressively.
We fit the composed shape to minimize the mean squared error (MSE) in
256 $x_\HI$'s and at 127 time points in each, and find the objective
value improve with $m$ but only marginally from $P_5$ to $P_7$.
Therefore, our final shape model is a composition of $\gomp$ and $P_5$,
(see the lower panel of Figure \ref{fig:shape}), and has 6 parameters to fit.

The complete model parametrizes the HI evolution as follows (Figure \ref{fig:shape}):
\begin{align}
\label{eq:uni}
x_\HI(\ar) &= \gomp\bigl( P_5(\ar) \bigr)
  \equiv \exp\bigl[ - \exp\bigl( P_5(\ar) \bigr) \bigr], \\
\label{eq:poly}
P_5(\ar) &= {\textstyle\sum}_{m=0}^5 \, c_m \ln^m\!\ar, \\
\mathbf{c} &= \{0, 1, 0.1130, 0.02600, 0.0005491, -0.00006518\}, \nonumber\\
\label{eq:map}
\ar(a; \vtheta) &= \Bigl[ \frac{a}{\ap(\vtheta)} \Bigr]^{\tilt(\vtheta)},
\end{align}
where $\vtheta$ denotes 6 astrophysical and cosmological parameters,
$\ap(\vtheta)$ is the power-law pivot (or logarithmic translation), and
$\tilt(\vtheta)$ is the rescaling tilt.
Their parameter dependences stem from \textsc{21cmFAST}'s modeling of
reionization astrophysics.

As for the transformation parameters, as in the PDF approach, we use an
affine transformation $\ln\ar = \tilt (\ln a - \ln\ap)$, or equivalently
a power law in Eq.~(\ref{eq:map}).
Because each $x_\HI$ has its own parameters of $\ap$ and $\tilt$, we
have in total $516 = 4 + 2 \times 256$ parameters to determine in the
joint fit.
$P_5$ only needs 4 instead of 6 parameters, because the constant and the
linear coefficients are fully degenerate with $\ln\ap$ and $\tilt$, and
therefore are fixed to be 0 and 1, respectively.
Figure \ref{fig:shape} shows all 256 $x_\HI$ timelines and their universality
after transformations.
We can then use the fitted $\ap$ and $\tilt$ as the target for symbolic
regression, to model their dependences on the independent cosmological
and astrophysical parameters.

Using the publicly available THESAN1 \cite{Kannan2022}
data\footnote{\url{https://www.thesan-project.com/data.html}}, we
demonstrate that more sophisticated simulations, which employ different
prescriptions for reionization astrophysics and incorporate radiative
transfer, also exhibit this universal behavior.
This consistency across various models is not unexpected, as most
simulations of the epoch of reionization converge on a similar overall
shape for the reionization timeline\footnote{This feature is currently
being exploited by the SKA collaboration in their Science data challenge
3, \url{https://sdc3.skao.int/challenges/inference}.}.
In Figure \ref{fig:thesan}, we show that our universal shape accurately fits
the THESAN1 simulation data. The comparison with the reionization timeline of the state-of-the-art THESAN simulation, alongside the material in Appendix \ref{app:halo}, provides supporting evidence that the universality may extend beyond the baseline \textsc{21cmFAST} setup. 

The early intergalactic medium is primarily composed by neutral hydrogen
and helium.
Neutral helium (HeI) loses its first electron at the same time as neutral
hydrogen (HI) gets ionized \cite{Trac2007}.
However, there is a second reionization that occurs around $z\sim3$
where Helium (HeII) loses its remaining electron.

CMB photons will scatter off any free electrons, therefore both helium
reionizations contribute to the Thomson optical depth to reionization
$\tau_\reio$, although HeII ionization contribute relatively little in
comparison to HI and HeI ionizations \cite{Liu2016}.

To include the impact of the first helium reionization in our Gompertz
\texttt{CLASS}, we assume it follows that of HI as done in the $\tanh$
model, i.e., the free electron fraction $x_\e$ is given by
\begin{align}
\label{eq:xe_H_He}
x_\e
&= \Bigl(1 + \frac{n_\He}{n_\mathrm{H}} - x^\rec_\e\Bigr) x_\e^\gomp
  + x^\rec_\e
\nonumber\\
&= \Bigl(1 + \frac{Y_\He}{C (1 - Y_\He)} - x^\rec_\e\Bigr) x_\e^\gomp
  + x^\rec_\e,
\end{align}
where $n_\He / n_\mathrm{H}$ is the helium to hydrogen number density
ratio, $C \equiv m_\He / m_\mathrm{H} \approx 4$ is their mass ratio,
$Y_\He$ is the helium mass fraction, $x_\e^\gomp$ corresponds to the
contribution of free electrons due to the Gompertzian contribution -- Eq.~(\ref{eq:uni}) -- and $x^\rec_\e
\approx 10^{-4}$ is the leftover free electrons from after
recombination.

Given the relatively small impact of the HeII reionization on
$\tau_\reio$, the current uncertainties regarding its timeline, and the
difficulty involved with its accurate modeling \cite{Hotinli2023,
Upton2020}, we opt to follow the conventional approach and include the
second Helium reionization using the $\tanh$ model
\begin{equation}
\label{eq:xe_tot}
x_\e^\mathrm{Tot} = x_\e + \frac{Y_\He}{2C(1 - Y_\He)}
  \biggl(\tanh{\Bigl(\frac{z_\re^\HeII - z}{\Delta z^\HeII}\Bigr)} + 1\biggr),
\end{equation}
where $z_\re^\HeII = 3.5$ and $\Delta z^\HeII = 0.5$ are the midpoint
and duration of the second helium reionization, respectively.
These choices are also the default values used by \texttt{CLASS}.

\subsection{Symbolic regression}
\label{sec:pysr}

\begin{figure}[tb]
\centering
\includegraphics[width=0.8\linewidth]{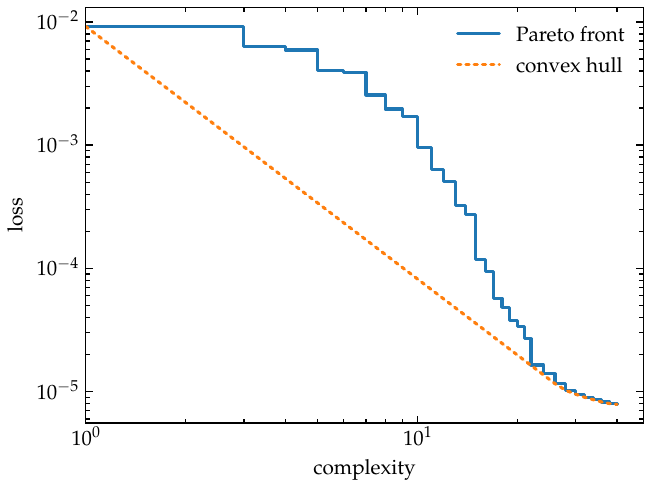}
\caption{\textbf{Pareto front for symbolic regression} (blue solid
steps) illustrates the trade-off between regression accuracy and
expression complexity.
We further add the lower convex hull to aid model selection: each
segment of the orange dotted line represents a power-law trade-off in
this log-log plot, and every model that touches the convex hull is more
economic -- in the sense of accuracy gain at cost of complexity -- than
the nearby models above the segments.
As an example, here we show the results of $\ln\ap(\vtheta)$, where the
complexity 22 point gives Eq.~(\ref{eq:SR_a}).}
\label{fig:pareto}
\end{figure}

SR learns a model of data in the form of analytic expressions.
Unlike traditional fittings that are restricted by their specific
parameterization, SR searches in the vast function space of all
expressions composed of specified operations, input variables, and free
constants.
It is NP-hard \cite{SongEtAl2024, VirgolinPissis2022} and typically need
genetic or deep learning-based algorithms.

To the authors' knowledge, the first application of SR in astronomy was
the rediscovery of classical astronomical relationships, such as the
fundamental plane of elliptical galaxies \cite{Graham2013}.
Since then, recent works have focused on using SR to uncover new
physical relations, for instance, identifying possible inflationary
potentials \cite{Sousa2024} and predicting $\Omegam$ from halo catalogs
\cite{Shao2023}.
For additional examples of SR-driven discoveries, interested readers are
directed to \url{https://astroautomata.com/PySR/papers/}.

In this work, we use the \texttt{PySR} package \cite{Cranmer2020b,
Cranmer2023} which performs SR optimizations using a multi-population
genetic algorithm and using the BFGS algorithm for optimizing constants
\cite{NocedalWright2006}.
With \texttt{PySR}, we search for symbolic expressions that take the 6
parameters in Figure \ref{fig:sobol} as inputs and output $\ln\ap$ or $\tilt$,
to minimize the MSE loss.
Here, the search space is the set of expressions composed of 5 binary
operators ($+$, $-$, $\cdot$, $/$, and power function) and 2 unary
operators ($\exp$ and $\ln$).
Each expression naturally takes the form of a binary tree, and the total
number of nodes is used as a coarse heuristic for the \emph{complexity}
of a symbolic expression.
Note that since we make a choice for the operators used in our search,
we are implicitly generating a prior over the space of expressions, such
that it looks similar to existing models described with these operators
-- as is much of physics.
We use 512 \texttt{PySR} populations each having 33 expressions, and
optimize for 20000 iterations each with 10000 cycles.

More complex expressions tend to fit more accurately, a trade-off
typically visualized by the Pareto front, as shown in Figure \ref{fig:pareto}.
Better and more economic expressions can achieve a lower loss at
moderate increase of complexity.
To aid model selection, we use a heuristic that compares all expressions
on the Pareto front globally, and only considers models that fare
favorably in power-law trade-offs of the form
\begin{equation}
\mathrm{loss}^{1 - \gamma} \cdot \mathrm{complexity}^\gamma
= \mathrm{const}, \quad \exists \gamma \in (0, 1).
\end{equation}
All such expressions lie on the lower convex hull\footnote{A related aid
for model selection uses the concept of hypervolume of the Pareto front
\cite{Cao2015}.} of the Pareto front, as illustrated in
Figure \ref{fig:pareto}.

For the pivot $\ln\ap$, we find complexity 22 is enough for the MSE loss
to reach $1.6 \times 10^{-5}$ (lower than sub-percent level error on
average), so there is no need to use expressions more complex than that.
For the tilt $\tilt$, complexity 25, corresponds to a loss $\approx
1.2\times10^{-3}$ ($\lesssim 0.05\%$ error on average), while higher
complexities can help but very slowly.
Therefore, based on the economic heuristic, we choose expressions of
complexity 22 and 25, for $\ln\ap$ and $\tilt$, respectively.

Our current selection for tilt and pivot could be a consequence of the
astrophysical assumptions of reionization in \texttt{21cmFAST}.
A more complex simulation, such as one involving radiative transfer or
different X-ray preheating \cite{Montero2024}, might reveal a different
mapping between physical parameters and $x_\HI$ profiles.

\subsection{Astrophysical data}
\label{ssec:xHI}
The increasing number of direct constraints on the reionization timeline
enhances CMB analyses.
Figure \ref{fig:history} includes upper limits from dark pixel
constraints \cite{Jin2023} (high-$z$ quasars), a lower bound from the
Ly$\alpha$ emission fraction \cite{Mesinger2015} (high-$z$ galaxies), and
an upper limit from the clustering of Ly$\alpha$
emitters \cite{Sobacchi2015} (high-$z$ galaxies).
It also presents constraints by Ly$\alpha$ equivalent width of
Ly$\alpha$ emitters \cite{Mason2018, Mason2019, Hoag2019} (high-$z$
galaxies) and quasar damping wings \cite{Greig2022, Greig2024, Spina2024,
Durovcikova2024} (high-$z$ quasars).
Furthermore, it includes the indirect constraints on $x_\HI$ through the
evolution of the galaxy Ly$\alpha$ luminosity
function \cite{Morales2021}.

In this work, we supplement Planck CMB data with quasar damping wing
(DW) constraints on $x_\HI$ (green pentagons in Figure \ref{fig:history}),
reflecting our confidence in their robustness.
Future studies could include luminosity function (LF) constraints as
more data becomes available, particularly since we find including them
could already halve the error on $\tau_\reio$, and because more data
will enhance our ability to constrain and marginalize over reionization
astrophysics.
We opt not to include the LF constraints on $x_\HI$ in this work as they
are in slight tension with the DW data at the current stage
(see Figure \ref{fig:history}).

While quasar damping wing provides a direct constraint on $x_\HI$ through its impact on the transmitted flux of the observed quasar, the Ly$\alpha$ LF indirectly infers $x_\HI$. This approach assumes that the declining fraction of Ly$\alpha$-emitting galaxies observed at $z\lesssim7$ is driven solely by the increasing $x_\HI$ in the IGM. However, this trend could also be due to evolving galaxy properties, like changes in the escape fraction of Ly$\alpha$ photons \cite{Dijkstra2014}. Moreover, constraints on $x_\HI$ from Ly$\alpha$ LF at relatively high redshifts ($z \sim 7$) exhibit tension with each other; for instance, \cite{2017ApJ...842L..22Z} reports $x_\HI \approx 0.4 - 0.6$, while \cite{2019ApJ...886...90H} finds $x_\HI \sim 0.2 - 0.4$. This discrepancy appears to be due to field-to-field variations in the number of observed Ly$\alpha$ emitters in a given volume arising from both large scale structure and the patchy nature of reionization \cite{2023ApJ...953...29B}. 

\section{Results}
\label{sec:res}
Before fully leveraging our methodology to extract the parameter
dependence in the rescaling of Equation~\eqref{eq:uni} and relaxing the need for
$\tau_\reio$ in CMB analyses, we first implement the Gompertzian shape with
independent $\tau_\reio$ in \textsc{CLASS} and confirm its agreement
with the conventional $\tanh$ model (gomp and $\tanh$ in
Table \ref{tab:uber-table}).
Using Planck PR3 likelihoods `TTTEEE' \cite{Planck2020c} and CMB lensing
\cite{Planck2020d}, we sample typical cosmological parameters with
\textsc{Cobaya} \cite{Torrado2020}, including $\tau_\reio$.
Given a proposal for $\tau_\reio$, we determine the corresponding
reionization timeline using bisection by varying $\ln\ap$ for gomp
while fixing $\tilt$ to its mean value,
meanwhile for $\tanh$, the reionization midpoint $z_\re$ is the tuning
parameter.
The sampler runs until the Gelman-Rubin statistic \cite{Gelman1992}
satisfies $R - 1 < 0.01$ for the variance in parameter means from
different chains.
We repeat this for $\tanh$ and verify the agreement between the two
models.

Figure \ref{fig:tg} and Table \ref{tab:uber-table}  summarize this
validation experiment.
The only notable differences in inferred parameters are in $z_\re$.
The gomp scenario suggests a more delayed reionization by over
$1\sigma$, with $z_\re = 6.76 \pm 0.67$ compared to $7.67 \pm 0.75$ for
$\tanh$, in alignment with recent high-$z$ quasar observations
\cite{Keating2020}.
All other cosmological parameters are in good agreement with Planck's
results \cite{Planck2020a}, with differences $\lessapprox 0.5 \%$.

Having demonstrated that the universally shaped reionization can
reproduce standard CMB analyses, we move to establish the connection
between the universal shape for $x_\HI$ and the rescaling of a given
reionization scenario.
We refer to this model as gomp + SRFull.
This rescaling naturally depends on cosmology and astrophysics.
For example, a larger matter density $\Omegam$ results in deeper
potential wells, accelerating structure formation and increasing the
number of ultraviolet photons driving the reionization process.
We employ \textsc{PySR}, an SR package, to establish the parameter
dependences of the rescaling in Eq.~\eqref{eq:map}.

While \textsc{PySR} initially guided us towards the Gompertz curve when
directly regressing $x_\HI$, the final analysis only uses it to regress
the pivot and tilt instead.
We fit their values jointly with the polynomial coefficients as
described above, and feed them as labels to the genetic algorithm to
find the best analytic expression (see \S\ref{sec:pysr} for our definition of \emph{best}).
Using \textsc{PySR} we derived the following mapping
\begin{align}
\label{eq:SR_a}
\ln\ap(\vtheta) &= \Bigl(\frac{\Omegab}{\Omegam}\Bigr)^h
  - (\sigma_8 - 0.04835) \bigl(\ns + 0.3558 \ln(0.1123 \zetaUV)\bigr) - \Omegam - \ns, \\
\label{eq:SR_b}
\tilt(\vtheta) &= \Bigl( \frac{0.005660^{\Omegam}}{0.6015}
    - \ln\bigl(\zetaUV - (\Omegam + \ns h)^{15.05}\bigr) + h \Bigr)
  \ln{\Omegab} + \frac{h}{\sigma_8},
\end{align}
where $\ns$, $h$, $\Omegab$, and $\Omegam$ are the tilt of the
primordial power spectrum, dimensionless Hubble constant, and present
baryon and matter density fractions, respectively.
$\sigma_8$ is the present linear rms relative density fluctuation in a
sphere of radius $8 \, h^{-1}$Mpc.

Eqs.~\eqref{eq:map}, \eqref{eq:SR_a}, and \eqref{eq:SR_b} are analytic expressions and thus
interpretable.
For example, higher values of $\Omegam$, $\sigma_8$, and $\zetaUV$
hasten reionization by enhancing structure formation and increasing the
abundance of ionizing photons.
Similarly, larger $\ns$ primarily expedites reionization by boosting
power on small scales, leading to more ionizing sources and earlier
completion \cite{Montero2021}.
Keep in mind that our \textsc{21cmFAST} simulations assume that faint
galaxies are the primary drivers of reionization.
Surprisingly, Eqs.~\eqref{eq:SR_a} and \eqref{eq:SR_b} suggests that higher $\Omegab$
delays reionization, likely due to more HI in the intergalactic medium
requiring additional ionizing photons.
The ratio $\Omegab/\Omegam$ in $\ap$ supports this view.
Moreover, $h$ exhibits competing effects: higher values increase physical densities
(potentially delaying reionization via $\omega_\mathrm{b} h^2$), they can also simultaneously enhance halo collapse fractions and therefore UV source formation. The latter effect dominates.

We note that within the prior range of our \textsc{21cmFAST} simulations
(see \S\ref{sec:sims}) and their
corresponding astrophysics of reionization, the mapping derived from SR
is not unique.
Additional details and results using an alternative mapping -- SRHalf --
are presented in Appendix \ref{app:SRHalf}.
Nonetheless, our results are robust and independent of the choice of
mapping.

We implement Eqs.~\eqref{eq:SR_a} and \eqref{eq:SR_b} in our Gompertz \textsc{CLASS},
which given the cosmological and astrophysical parameters, determines
the pivot and tilt values, and consequently the reionization history,
$\tau_\reio$, and CMB angular power spectra.
This gomp + SRFull model eliminates the need to sample over $\tau_\reio$
(or $z_\re$), requiring only five cosmological parameters and $\zetaUV$.
Moreover, thanks to SR, the model can constrain $\zetaUV$ using CMB and
astrophysical data directly, a link that was not utilized in
previous efforts \cite{Greig2017}.
The advantage of sampling over $\zetaUV$ instead of $\tau_\reio$ is that
powerful astrophysical constraints on $x_\HI$ can be used to marginalize
over $\zetaUV$ -- note that a more complex parameterization of 
astrophysical parameters, beyond a single constant $\zetaUV$, can 
still be marginalized over following the same method.
We use \textsc{Cobaya} to re-analyze the same CMB data and include
astrophysical data (see \S\ref{ssec:xHI}) to
effectively constrain and marginalize over astrophysics ($\zetaUV$).

\begin{figure*}[!htb]
\centering
\includegraphics[width=0.8\linewidth]{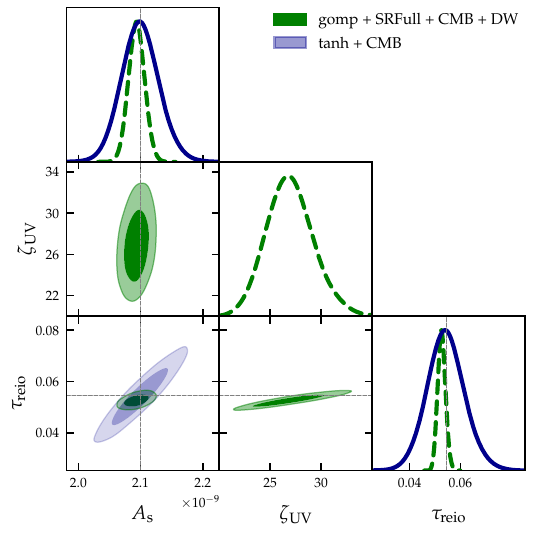}
\caption{Analysis of CMB and astrophysical data with
reionization as a function of cosmology and astrophysics.
The green contours represent our results using the Gompertzian reionization
model with Eqs.~\eqref{eq:SR_a} and \eqref{eq:SR_b}, which eliminates the need to sample
over any conventional reionization parameter and uses quasar damping
wing data on $x_\HI$ to constrain and marginalize over reionization
astrophysics.
The blue contours correspond to the results obtained using the
conventional $\tanh$ model, while the relevant Planck constraints
\cite{Planck2020a} are depicted with gray lines for reference.
The $\tanh$ model performs poorly in fitting astrophysical data; thus,
combining $\tanh$ with astrophysical data is ill-advised.}
\label{fig:kill}
\end{figure*}

Figure \ref{fig:kill} underscores the impact of our universally-shaped
Gompertzian reionization model, tightening the optical depth constraint to
$<3\%$ compared to $> 10\%$ with the $\tanh$ prescription (see Appendix \ref{app:vali} for a comparison when using only the Gompertzian shape and sampling over $\tau_\reio$).
Furthermore, the constraint on $\As$ improves dramatically since the TT
data is no longer significantly hampered by the degeneracy between $\As$
and $\tau_\reio$.
The error on $\As$ decreases by an impressive factor of 2.3 compared to
Planck's results \cite{Planck2020a}.
Overall, we recover tighter constraints across the board compared to
Planck.
Notably, combining SR with CMB and astrophysical data, specifically
quasar damping wing (DW) data, yields a highly competitive constraint on
$\zetaUV = 26.9^{+2.1}_{-2.5}$, a commanding improvement over previous direct
constraints (e.g., $\zetaUV = 28^{+52}_{-18}$) \cite{Greig2017}.
See Appendix \ref{app:mcmc}, Figure \ref{fig:unleashed_gomp}, and Table \ref{tab:uber-table}.

\section{Summary and discussion}
\label{sec:sum}
\begin{figure*}[!htb]
\centering
\includegraphics[width=0.8\linewidth]{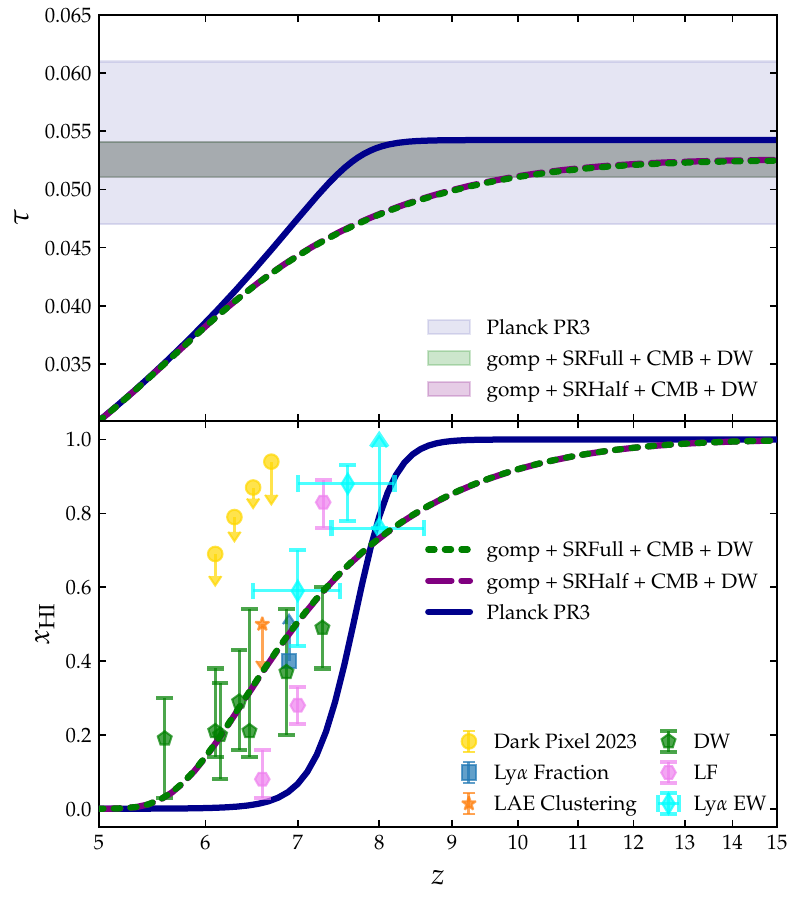}
\caption{Optical depth evolution $\tau(z)$ and reionization
history $x_\HI(z)$.
Our best-fit gomp + SRFull model (green dotted line) of Planck CMB data
+ quasar damping wing (DW) data is asymmetric and differs significantly
from that of the symmetric $\tanh$ model (blue solid line).
Note $z$ is shown in logarithmic scale of $a^{-1}$.
We also include an alternative mapping from physical parameters to
Gompertz timeline -- gomp + SRHalf -- in the purple dashed line (see Appendix \ref{app:SRHalf}).
The shaded regions in the upper panel correspond to the inferred range
in $\tau_\reio$ from analyzing Planck PR3 data.
Additionally, the lower panel includes observational constraints from
high-redshift quasars and galaxies (see \S\ref{ssec:xHI}).}
\label{fig:history} 
\end{figure*}

Our results suggest that Planck data favors a delayed reionization
compared to other CMB-based constraints (in a $\chi^2$-sense).
Our best-fit cosmological parameters indicate a midpoint of $z_\re =
6.98$ and a duration of $\Delta z \equiv z(x_\HI = 0.05) - z(x_\HI =
0.95) \approx 560 $ Myr, consistent with our previous constraint
without SR.
While our results align with late reionization observations, the
difference from $\tanh$ is within 1$\sigma$.
The duration of reionization, though better suited to observational
constraints compared to $\tanh$, might still be considered somewhat
rapid in the context of late reionization scenarios \cite{Cain2021}.
Figure \ref{fig:history} illustrates the reionization timeline derived from
our best-fit values, in contrast with the poor fit of symmetric $\tanh$
to the astrophysical data.

Our findings for the timeline of reionization align with late
reionization scenarios, which are supported by high-$z$ Lyman-$\alpha$
observations \cite{Keating2020, Cain2021}.
However, recent discoveries by JWST indicate the presence of massive,
bright galaxies at early redshifts $z \sim 10$ \cite{Adams2023,
Bradley2023, Donnan2023}.
The presence of these early galaxies suggests a potential preference for
brighter galaxies to drive reionization, a role that in our
\textsc{21cmFAST} simulations was attributed to a population of faint
galaxies.

We note that our results assume a flat Universe.
Relaxing this assumption requires adding an extra parameter to the
functional form of $\tau_\reio$ \cite{Anselmi2023}.
Furthermore, our results are influenced by the semi-numerical
prescription employed by \textsc{21cmFAST} to ionize the IGM, which,
while efficient and swift, could bias our findings.
Moreover, our exploration within the astrophysical framework of
\textsc{21cmFAST} has been limited to varying the ionization efficiency
(see \S\ref{sec:sims}). Thus, this work has eliminated the need to sample over $\tau_\reio$ conditional on the adopted reionization framework.
A valuable test to validate our reionization history involves using
simulations with accurate radiation transport and realistic galaxy
formation physics.
One state-of-the-art option is the THESAN simulations \cite{Kannan2022},
which also display the universal $x_\HI$ shape, as shown in
Figure \ref{fig:thesan}. Consequently, given enough computational resources, one could reproduce the elimination of $\tau_\reio$ using the THESAN simulations and the framework introduced in this work.
Additionally, a more comprehensive exploration is warranted to ensure,
for instance, that our findings for $\ap$ and $\tilt$ will not bias
cosmological analyses when incorporating additional astrophysical
parameters, such as the minimum halo mass that host ionizing sources. Ref. \cite{2026arXiv260413423M} demonstrates that there is little impact on our results when accounting for a more robust reionization model. 

Throughout this work we considered only quasar damping wings for astrophysical data,
but including current luminosity function constraints can already halve
the error on $\tau_\reio$.
Hence, as reionization constraints improve, we anticipate substantial
cosmological gains with our framework.

\appendix

\section{Universality in a halo-mass dependent ionization efficiency}
\label{app:halo}
In \S\ref{sec:uni}, we obtained a universality for the evolution of the neutral hydrogen fraction under the assumption of a constant ionization efficiency. Here, we relax this assumption by employing the reionization prescription introduced in \cite{Park2019} while simultaneously varying the underlying cosmology\footnote{We follow the \emph{core} sampling detailed in \S\ref{sec:sims}.}. In this parameterization, $\zetaUV$ is replaced with $f_*$ and $f_{\rm esc}$ (the escape and galactic gas in stars fractions for $10^{10}$ M$_\odot$ haloes, respectively) and their corresponding power-law indices, $\alpha_*$ and $\alpha_{\rm esc}$, which capture the halo mass dependence. Furthermore, we account for additional astrophysical feedback and heating through the threshold mass for quenching of star formation in haloes ($M_{\rm turn}$), the fractional characteristic time-scale defining the star-formation rate of galaxies ($\tau_*$), and the X-ray parameters $E_0$ and $L_{\rm X}$, which define the X-ray energy threshold for self-absorption by host galaxies and the specific X-ray luminosity of early X-ray sources, respectively.

We draw 42 $x_\HI$ samples from the core cosmology range alongside the astrophysical parameter ranges listed below:

\begin{alignat}{4}
\label{eq:prior_app}
\log_{10}(f_*) &\in (-2, -1), &\quad
\alpha_* &\in (0.3, 0.7), &\quad
\tau_* &\in (0.2, 0.6), &\quad
\log_{10}(M_{\rm turn}) &\in (8., 9.), \nonumber\\
\log_{10}(f_{\rm esc}) &\in (-2, -0.7), &\quad
\alpha_{\rm esc} &\in (-1, -0.2), &\quad
E_0 &\in (300, 1200), &\quad
\log_{10}(L_{\rm X}) &\in (39, 41).
\end{alignat}

\begin{figure}
    \centering
    \includegraphics[width=0.7\linewidth]{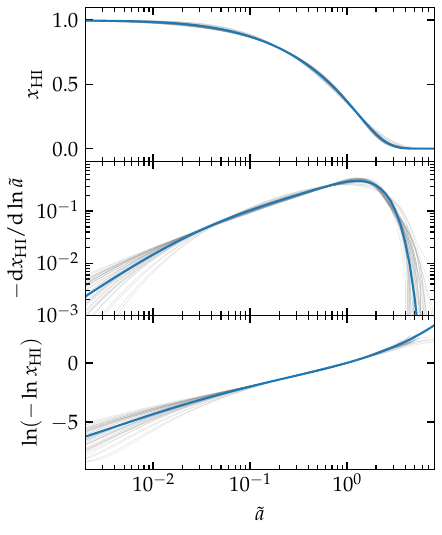}
    \caption{\textbf{\boldmath Universality holds for non-constant $\zetaUV$.} Same as Figure \ref{fig:shape} but for 42 $x_\HI$ samples derived utilizing the halo-mass dependent reionization prescription introduced in \cite{Park2019}. In addition to mass-dependent ionization efficiency, we vary astrophysical parameters that govern the difficulty of forming the formation efficiency of ionizing sources and the impact of X-ray preheating.}
    \label{fig:halo}
\end{figure}

Figure \ref{fig:halo} demonstrates that the universality is preserved under a mass-dependent prescription for the ionization efficiency. Note that to compute Figure \ref{fig:halo} we have recalibrated the $P_5(\tilde{a})$ coefficients in Eq.~(\ref{eq:poly}) using the 42 new $x_\HI$ samples.

\section{Validation of the universal shape with MCMC}
\label{app:vali}
\begin{figure}
\centering
\includegraphics[width=\linewidth]{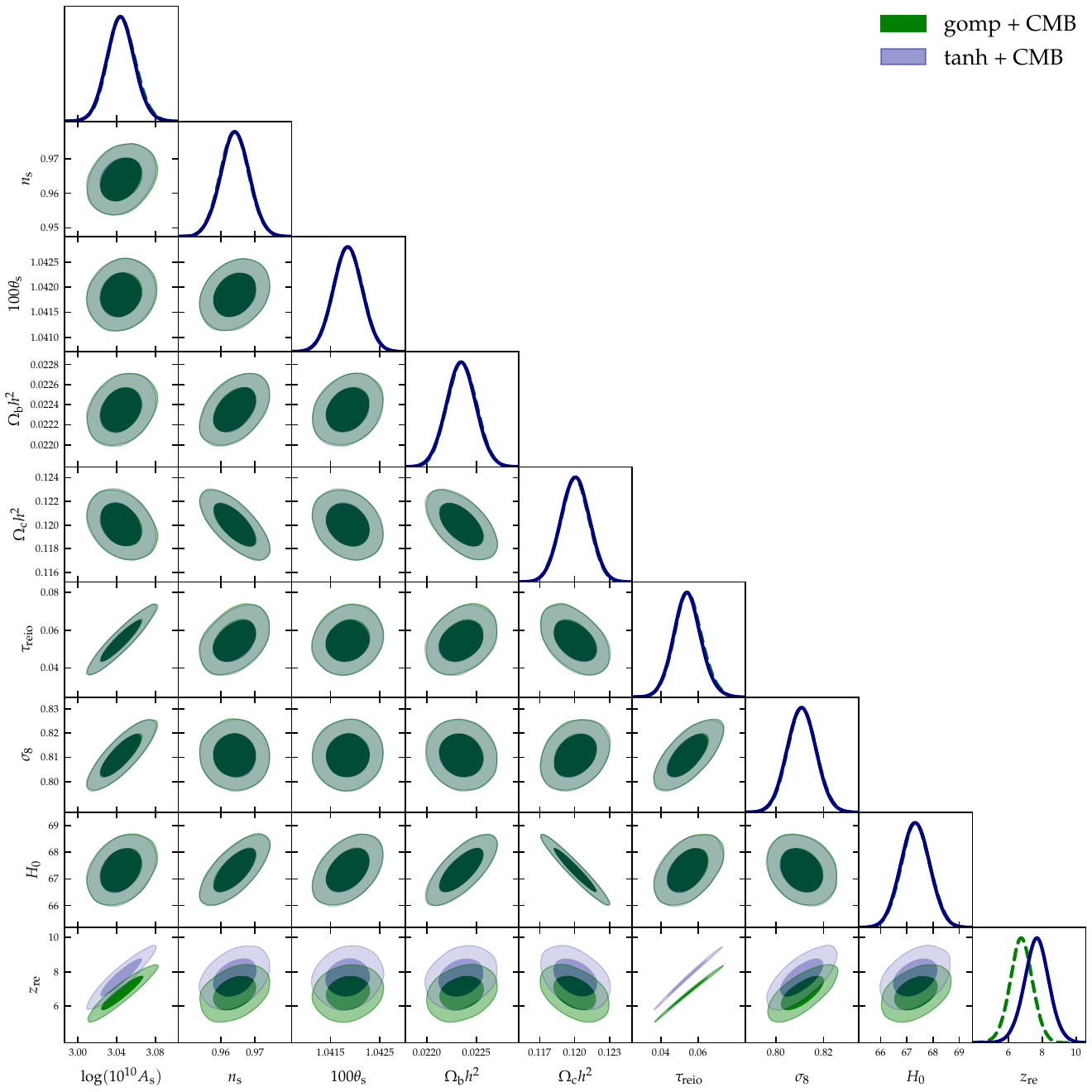}
\caption{\textbf{\boldmath Validation of our universal shape for $x_\HI$
in Eqs.~(\ref{eq:uni}) and (\ref{eq:poly}) vs.\ the conventional $\tanh$ model using CMB
data with sampling over $\tau_\reio$.} 
Note that we have not use our mapping from physical parameters to
$\tau_\reio$.
Here, we use the standard choice of sampling over optical depth and
obtain the corresponding reionization history via bisection.
The green (blue) contours correspond to the constraints obtained with
our gomp ($\tanh$) model.}
\label{fig:tg}
\end{figure}

Figure \ref{fig:tg} showcases the result of an inference of CMB data that neglects the symbolic regression part of our work, i.e., it only uses the universal shape for the timeline of reionization instead of the conventional hyperbolic tangent prescription. 

As shown in Figure \ref{fig:tg}, if one ignores the mapping between reionization and cosmology, the Gompertzian reionization model and the conventional hyperbolic tangent prescription agree extremely well with each other for all cosmological parameters, including $\tau_{\reio}$ which is being sampled\footnote{We use bisection to obtain $\alpha$ and fix $\beta$ to a constant value of 7.66 based on the complexity 1 expression from \texttt{PySR}.}. See Table \ref{tab:uber-table} for the specific values. Unsurprisingly, the corresponding midpoint of reionization differs between both models since the Gompertzian model allows for a later reionization scenario compared to that of $\tanh$.

\section{Alternative mapping}
\label{app:SRHalf}
The mapping between neutral hydrogen profiles and cosmology is not
unique.
This is because symbolic regression algorithms can be non-convergent.
Moreover, the resulting symbolic expressions can depend on the data used
for training, i.e., overfitting.

To investigate the impact of overfitting, we train an alternative
mapping of reionization with cosmology using only half of the
\texttt{21cmFAST} $x_\HI$ profiles, which we refer as SRHalf.
For the pivot and tilt, we obtain
\begin{align}
\label{eq:SRHalf_a}
\ln\ap(\vtheta) &= \Bigl(\frac{\Omegab}{\Omegam}\Bigr)^{\Omegam}
  - \ln^{0.5271} \Bigl(h \bigl(\zetaUV
    + \Omegab^{-0.4982}\bigr)^{\sigma_8}\Bigr)
  - \ns^{1.834}, \\
\label{eq:SRHalf_b}
\tilt(\vtheta) &= \biggl(\frac{\zetaUV - \Omegam^{-1.583}}{\Omegab h}
  \biggr)^{0.3163},
\end{align}
with a training loss (complexity) of $1.1 \times 10^{-5}$ (22) and $3.7
\times 10^{-3}$ (11), respectively.
Overall, we obtain a relative root mean squared error of 0.9\% when
reproducing the simulated optical depths from the $x_\HI$ (edge + core)
with SRHalf.
Now we can test this on the other half of our $x_\HI$ sample and get a
validation loss of $2.2 \times 10^{-5}$ and $6.5 \times 10^{-3}$
respectively, only slightly larger than the training loss on absolute
scales, which implies that we are safe from overfitting, supported by
consistency of their MCMC results.

Unsurprisingly, Eqs.~(\ref{eq:SRHalf_a}) and (\ref{eq:SRHalf_b}) recover the cosmological
trends discovered with SRFull.
Specifically, increasing $\zetaUV$, $\ns$, $\Omegam$, and $\sigma_8$
primarily leads to earlier reionization, while larger $\Omegab$ delays
it.
The ratio of matter densities appears in the pivot expression again.
Notably, the role of $h$ generally shows the opposite trend in $\ap$.
We note that these parameter trends are also generally present in lower
complexity expressions in the Pareto front.
Furthermore, Eq.~(\ref{eq:SRHalf_b}) excludes $\ns$ and $\sigma_8$ because of
our choice of complexity for SRHalf, which, despite having a larger
loss, still achieves mean squared errors comparable to SRFull when
reproducing the simulated data.

Replacing Eqs.~(\ref{eq:SR_a}) and (\ref{eq:SR_b}) with Eqs.~(\ref{eq:SRHalf_a}) and (\ref{eq:SRHalf_b}), we
rerun the analysis following the same approach, and summarize the
results using CMB jointly with astrophysical data in
Table~\ref{tab:uber-table}.
Comparing these findings to our previous results, we observe strikingly
similar values for all cosmological parameters.
The consistency between our results using the full $x_\HI$ sample and
those using only half of it indicates that our symbolic expression
mappings perform robustly and are not biasing the parameter inferences.

\begin{figure}
\centering
\includegraphics[width=\linewidth]{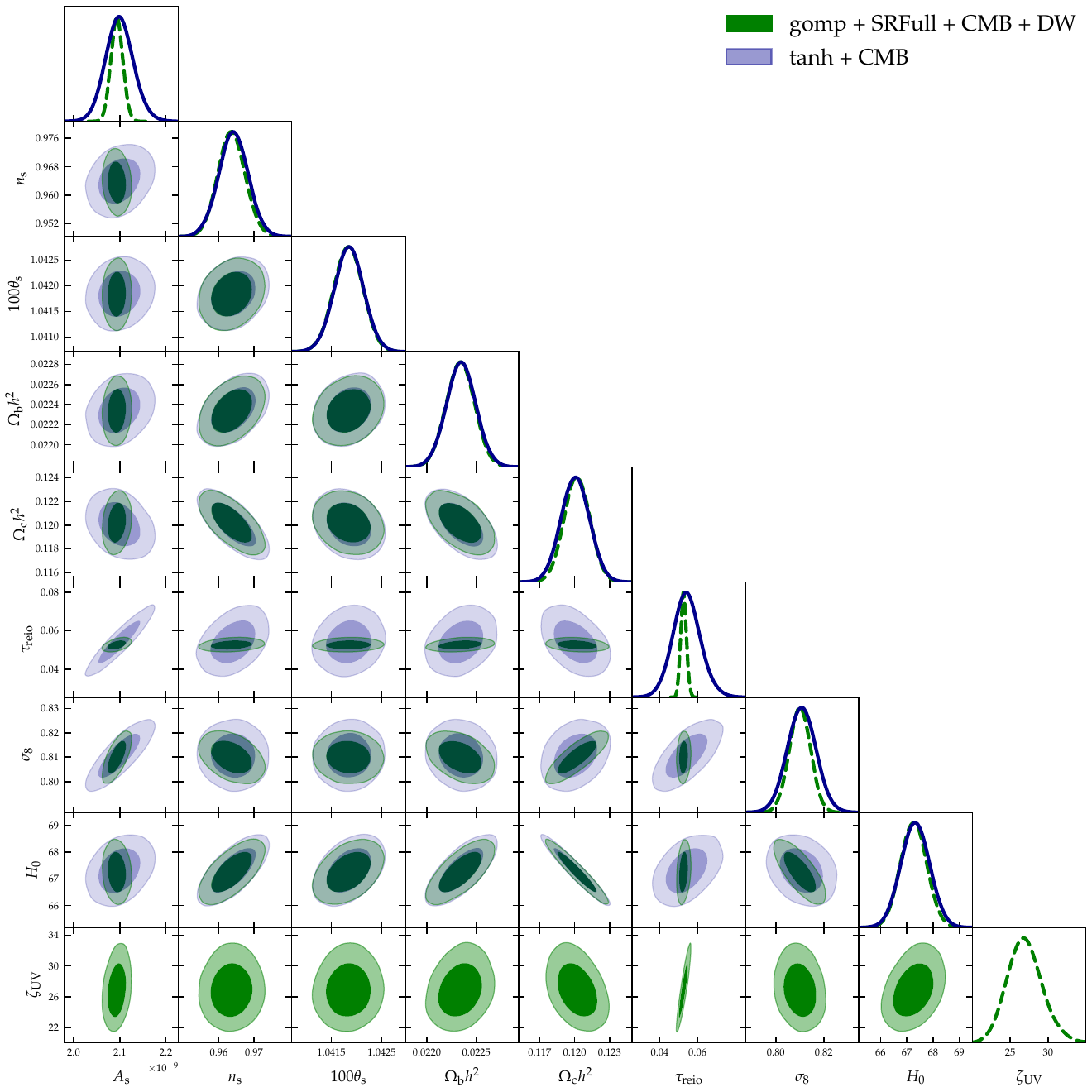}
\caption{\textbf{\boldmath Analysis of CMB data and astrophysical data
treating $\tau_\reio$ as a derived parameter using
Eqs.~(\ref{eq:uni}) to (\ref{eq:SR_b}) vs.\ sampling it with the
conventional $\tanh$ model.}
Here gomp + SRFull combines our Gompertzian universal shape with the
rescaling pivot and tilt obtained via symbolic regression.
Note that for gomp + SRFull we do not sample over $z_\re$, since
Eq.~(\ref{eq:uni}) does not depend on it but we use the astrophysical data to
constrain and marginalize over reionization astrophysics.
The green (blue) contours correspond to the constraints obtained with
our Gompertzian universal shape ($\tanh$ model).}
\label{fig:unleashed_gomp}
\end{figure}

\begin{table}[!t]
\centering
\caption{\textbf{Parameter constraints from Planck CMB and astrophysical
(DW) data.}
Summary table of the constraints for representative parameters of the
universal shape (Gompertz) and $\tanh$ reionization models.
The constraints use CMB `TTTEEE' + lensing information alone or jointly
with quasar damping wing $x_\HI$ constraints.
The results from Planck analysis \cite{Planck2020a} are included for
comparison.
The validation models are gomp and $\tanh$, gomp does not include the
cosmological and astrophysical dependences present in the timeline
of reionization, see also Appendix \ref{app:vali} for a visualization of these models.
In contrast, gomp + SRFull, and the alternative mapping gomp + SRHalf,
both include the additional parameter dependences in the reionization
timeline.
The shaded cells highlight the parameters that are sampled over by MCMC
for the different models.
The numbers in parentheses give the marginalized 1$\sigma$ uncertainty
in the last two significant digits.
We highlight the best constraints in boldface, with improvements by
factors of 5 and 2.3 on $\tau_\reio$ and $A_s$, respectively.
We also highlight our extremely competitive $\zetaUV$ constraint.
The age of the Universe and $H_0$ are in unit of Gyr and km/s/Mpc,
respectively, and $S_8 \equiv \sigma_8 \sqrt{\Omegam/0.3}$ as usual.}
\makebox[\textwidth][c]{\small
\renewcommand{\arraystretch}{1.3}
\begin{tabular}{c *{3}{r} !{\hspace{.5em}} *{2}{r} }
\toprule
 & \multicolumn{3}{c}{CMB} & \multicolumn{2}{c}{CMB + DW} \\
\cmidrule(lr){2-4} \cmidrule(lr){5-6} 
Parameter & Planck PR3\cite{Planck2020a} & $\tanh$ & gomp & gomp + SRFull & gomp + SRHalf \\
\midrule
$10^{9} \As$ & \sampled 2.100(30) & \sampled 2.099(30) & \sampled 2.101(31) & \textbf{2.094(13)} & \textbf{2.094(13)} \\
$\ns$ & \sampled 0.9649(42) & \sampled 0.9641(42) & \sampled 0.9640(42) & \sampled 0.9636(39) & \sampled 0.9637(39) \\
$\Omegac h^2$ & \sampled 0.1200(12) & \sampled 0.1201(12) & \sampled 0.1200(12) & \sampled 0.1202(11) & \sampled 0.1202(11) \\
$\Omegab h^2$ & \sampled 0.02237(15) & \sampled 0.02235(15) & \sampled 0.02235(15) & \sampled 0.02234(14) & \sampled 0.02234(14) \\
$\Omegam$ & 0.3153(73) & 0.3157(76) & 0.3155(75) & 0.3166(69) & 0.3163(69) \\
$\Omegam h^2$ & 0.1430(11) & 0.1430(11) & 0.1430(11) & 0.1432(10) & 0.1431(10) ) \\
$\OmegaL$ & 0.6847(73) & 0.6842(76) & 0.6845(75) & 0.6833(69) & 0.6836(69) \\
Age & 13.797(23) & 13.800(23) & 13.799(23) & 13.802(22) & 13.801(22) \\
$H_0$ & 67.36(54) & 67.32(55) & 67.34(55) & \sampled 67.26(50) & \sampled 67.28(50) \\
$100 \theta_\mathrm{X}$ & \sampled 1.04092(31) & \sampled 1.04185(29) & \sampled 1.04185(29) & 1.04184(29) & 1.04184(29) \\
$\sigma_8$ & 0.8111(60) & 0.8108(60) & 0.8109(60) & \sampled 0.8100(44) & \sampled 0.8099(44) \\
$S_8$ & 0.832(13) & 0.832(13) & 0.832(13) & 0.832(13) & 0.832(13) \\
$\tau_\reio$ & \sampled 0.0544(73) & \sampled 0.0543(75) & \sampled 0.0547(76) & \textbf{0.0526(15)} & $\mathbf{0.0527^{+(14)}_{-(16)}}$ \\
$z_\re$ & 7.67(73) & 7.67(75) & 6.76(67) & 6.98(17) & 6.98(17) \\
$\zetaUV$ & & & & \sampled $\mathbf{26.9^{+2.1}_{-2.5}}$ & \sampled $\mathbf{27.0^{+2.1}_{-2.5}}$ \\
\bottomrule
\end{tabular}}
\label{tab:uber-table}
\end{table}

\section{MCMC inference}
\label{app:mcmc}
Our inference combines CMB with astrophysical data.
Table~\ref{tab:uber-table} summarizes the results obtained by performing MCMC
Bayesian inference with \texttt{Cobaya} for the models considered
throughout this work.
We also include the Planck results \cite{Planck2020a} for reference.
In total, we run one Gompertz and one $\tanh$ reionization models with
CMB data and two extra Gompertz models using SR and astrophysical data
jointly with CMB to infer cosmology while constraining and
marginalizing over $\zetaUV$.

When considering only CMB data, we sample the typical 6 cosmological
parameters (see shaded parameters in Table~\ref{tab:uber-table} and Appendix \ref{app:vali}), sampling
$\tau_\reio$ and using bisection to obtain the reionization timeline.
For $\tanh$, the bisection varies $z_\re$, while for Gompertz we fix
$\tilt$ to the mean value (7.66, the complexity 1 expression from
\texttt{PySR}) and vary $\ap$.
Therefore, we do not use the SR mapping from physical parameters to
reionization -- Eqs.~(\ref{eq:SR_a}) and (\ref{eq:SR_b})  or alternatively
Eqs.~(\ref{eq:SRHalf_a}) and (\ref{eq:SRHalf_b}).

In contrast, when astrophysical data is included, $\tanh$ samples over
$z_\re$ instead of $\tau_\reio$ and an additional likelihood component
is calculated in \texttt{Cobaya}.
For Gompertz, we do not sample any reionization parameter.
Instead, we use SRFull or SRHalf to map physical parameters to
reionization timeline and include the astrophysical likelihood
component.
Perhaps unsurprisingly, $\tanh$ reionization does a poor job of fitting
the astrophysical data, with a $\chi^2$ larger than Gompertz's by an
order of magnitude, hence we do not show the $\tanh$ + CMB + DW results
in Table~\ref{tab:uber-table}.

There is an apparent discrepancy between the different models in
$\theta_\mathrm{x}$, a proxy for the angular scale of the acoustic
oscillations $\theta_*$.
The difference is due to our use of $100\theta_\mathrm{s}$, which
corresponds to the peak scale parameter defined exactly as the ratio of
the sound horizon divided by the angular diameter at decoupling, with
decoupling time given by the maximum of the visibility function, i.e.,
the standard choice for CLASS.
In contrast, the Planck collaboration reports $100\theta_\mathrm{MC}$,
which is given by Eq.~(6) in \cite{Planck2014}, and corresponds to the
standard choice in \texttt{CosmoMC}\cite{Lewis2002}.
Unsurprisingly, different reionization scenarios give similar values of
$100\theta_\mathrm{s}$.

Although Table~\ref{tab:uber-table} does  display an error in the midpoint
of reionization for gomp + SRFull + CMB + DW and gomp + SRHalf + CMB +
DW models, it is important to note that the error has been obtained by post-processing the chains.
Unfortunately, it cannot be automatically computed with \texttt{Cobaya}
because SRFull and SRHalf models do not sample over reionization
parameters.
Therefore, $z_\re$ has no meaning in our Gompertz
\texttt{CLASS} for models that take advantage of
the mapping between physical parameters and reionization timeline.

Remarkably, the inferred value of $S_8$ remains consistent across all
reionization models, even when $\sigma_8$ and $\Omegam$ show small
variations between models.
This trend suggests that different reionization models neither alleviate
nor exacerbate the $S_8$ tension.
While this observation holds for the Planck PR3 data and for CMB + DW,
we anticipate that incorporating low-redshift Baryon Acoustic
Oscillation (BAO) data may affect this conclusion.
Ref. \cite{2026arXiv260413423M} explores the implications of Gompertzian reionization for
the joint analysis of CMB and low-redshift data.

\acknowledgments

We thank David Spergel and Francisco Villaescusa-Navarro for comments that helped improve the manuscript. This work is supported by the Basic and Frontier Research Project of PCL (grant No.~2025QYB012) and the Major Key project of Peng Cheng Laboratory. The authors acknowledge PCL's Cloud Brain for providing computational and data storage resources that have contributed to the results reported within this paper.

\section*{Code \& Data Availability}
The source files, code scripts, and data results in this work are available in our GitHub repository {\sc 5par} (\url{https://github.com/eelregit/5par}). Our fork of \textsc{CLASS} with Gompertzian reionization is available in our repository {\sc class\_gomp} (\url{https://github.com/paulomontero/class_gomp}).


\bibliographystyle{JHEP}
\bibliography{biblio.bib}


\end{document}